\documentclass[12pt,preprint]{aastex}
\begin{document}
\title{Sources for Metals in the Intergalactic Medium}
\author{Y.-Z. Qian\altaffilmark{1} and G. J. Wasserburg\altaffilmark{2}}
\altaffiltext{1}{School of Physics and Astronomy, University of
Minnesota, Minneapolis, MN 55455; qian@physics.umn.edu.}
\altaffiltext{2}{The Lunatic Asylum, Division of Geological and
Planetary Sciences, California Institute of Technology, Pasadena,
CA 91125.}

\begin{abstract}
We present a discussion of possible sources of C, O, and Si in the
intergalactic medium (IGM) using the yields of very massive stars
(VMSs) and Type II supernovae (SNe II).
The chemical evolution of the IGM is considered
based on analytical phenomenological models of hierarchical
structure formation. Two regimes are considered: one for 
gas expulsion by VMSs and SNe II in
low-mass halos prior to dissociation of H$_2$ molecules and
reionization, and the other for later SN II-driven outflows from
intermediate-mass halos. We use recent data on the abundances
of C, O, and Si 
in the IGM inferred from two UV background (UVB) models.
We show that the results from a
UVB model including quasars only cannot be explained by 
existing stellar models. To account for the
results, in particular [Si/C]$_{\rm IGM}$,
from a softer UVB model requires 
VMSs to provide $\gtrsim 15$\% of the C in the IGM. The
preferred scenario is that VMSs in low-mass halos provided 
between 15\% and 60\% of the C
in the regime of very high 
redshift ($z\gtrsim 15$) and that galactic outflows provided
the remainder during later epochs 
($4<z\lesssim 6$). Thus, there is a large gap in $z$
between metal production by VMSs very early in the chemical evolution 
of the IGM and subsequent contributions from galactic outflows.
The observational estimate of [O/H]$_{\rm IGM}$ implies a high
[Fe/H]$_{\rm IGM}\geq -3$ for all cases considered 
(including the case of a pure VMS source). This raises problems with 
regard to observations of metal-poor stars in the Galaxy. 
In addition, the [O/Fe] values for SN II models are in conflict with
stellar observations, which indicate that the calculated average
Fe yields of SNe II are a factor of $\sim 2$ too high.
These issues and the general problem of relating abundances in the 
IGM to those in metal-poor stars remain to be investigated.
\end{abstract}


\section{Introduction}
We present a discussion of possible sources of C, O, and Si
in the intergalactic medium (IGM) with particular consideration
of very massive ($>100\,M_\odot$) stars (VMSs) and Type II
supernovae (SNe II).
It is now well established that there is a substantial inventory of metals
in the general IGM over a wide range of redshift $z=1.5$--5.5
(e.g., Songaila 2001; Pettini et al. 2003). Detailed studies of the data
over $z=1.8$--4.1 by Schaye et al. (2003) and Aguirre et al. (2004)
show that at a given density the abundances of C and Si follow
a lognormal distribution with substantial scatter.
Both the median and the scatter of the distribution
are dependent on the density but nearly independent of $z$. Using
observations at $z\sim 2.5$ Simcoe, Sargent, \& Rauch (2004) also found
lognormal distributions for C and O. As only the ions \ion{C}{4},
\ion{O}{6}, and \ion{Si}{4} are directly measured,
the inferred elemental abundances and in particular the relative
abundances of C, O, and Si are sensitive to the model of UV background
(UVB) used (e.g., Giroux \& Shull 1997). 
We will use the results for two common UVB models adopted by
the above two groups (see Table 1). For UVB model Q including quasars only,
Schaye et al. (2003) found the
average C abundance in the IGM to be [C/H]$_{\rm IGM}\approx -2.3$
over the density range
of the data used ($\log\delta=-0.5$ to 1.8 with $\delta$ being
the density relative to the cosmic mean) and up to 
[C/H]$_{\rm IGM}\approx -2.0$ when
extrapolated to higher densities. The corresponding [Si/C]$_{\rm IGM}$ is
$1.45$ (Aguirre et al. 2004). There is no evidence for evolution of either
[C/H]$_{\rm IGM}$ or [Si/C]$_{\rm IGM}$ over $z=1.8$--4.1 covered by the
data. For a similar UVB model, Simcoe et al. (2004) found
[C/O]$_{\rm IGM}=0$ and their lognormal distribution corresponds to
[C/H]$_{\rm IGM}=-2.2$.
For a different UVB model QG including both quasars and galaxies,
Schaye et al. (2003) and Aguirre et al. (2004) found
[C/H]$_{\rm IGM}=-2.8$ to $-2.5$ and [Si/C]$_{\rm IGM}=0.74$ while
Simcoe et al. (2004) found [C/O]$_{\rm IGM}=-0.5$ and a median abundance
of [C/H]~$=-3.1$, which would correspond to [C/H]$_{\rm IGM}=-2.5$ for a
lognormal distribution with the same scatter of 0.75 dex as for model Q.
For both models Q and QG, the results of the two groups for
[C/H]$_{\rm IGM}$ are in good agreement. Note that model Q gives higher
values for [C/H]$_{\rm IGM}$ and much higher values for [C/O]$_{\rm IGM}$
and [Si/C]$_{\rm IGM}$ than model QG (see Table 1).
Considering uncertainties in both
the measurements and UVB models, the average abundances in the IGM are
rather high with [O/H]$_{\rm IGM}=-2.3$ to $-2.0$. The C/O and Si/C ratios 
in the IGM do not appear to vary with density
(Simcoe et al. 2004; Aguirre et al. 2004). Both [C/H]$_{\rm IGM}$ and 
[Si/C]$_{\rm IGM}$ appear to be
independent of $z$ over at least the range of $z=1.8$--4.1.

An important question is when the metal inventory of the IGM was
provided. At present we know that this inventory was achieved
prior to $z\sim 4$. Metal ejection from both low-mass ($\sim
10^5$--$10^7\,M_\odot$) halos at early epochs ($z\gtrsim 15$) and
intermediate-mass ($\sim 10^8$--$10^{10}\,M_\odot$) halos at later
epochs ($z<15$) will be discussed. The stellar sources of these
metals are also not known but VMSs and SNe II are assumed to be
the plausible candidates. We earlier proposed a prompt metal
inventory from VMSs (e.g., Wasserburg \& Qian 2000). Using
estimates of this inventory and yields of Heger \& Woosley [2002
(HW)], Oh et al. (2001) suggested that VMSs could be the source of
reionization (cf., Venkatesan \& Truran 2003). 
It will be shown that if the IGM inventory was
obtained at early epochs, the required metal production rate for
VMSs and zero-metallicity SNe II in low-mass halos is much higher
than the Galactic rate. In contrast, the bulk of the IGM inventory
can be provided at later epochs by low-metallicity SNe II in
intermediate-mass halos through efficient outflows for a Galactic
metal production rate. However, in all cases and for all the
stellar models considered there is a problem in accounting for the
elemental ratios inferred for the IGM from UVB model Q. It will be
shown that the ratios from model QG can be matched with
significant contributions from VMSs. In particular, the Si/C ratio
reported for the IGM appears to require a blend of contributions 
from both VMSs and SNe II if Salpeter initial mass functions (IMFs) 
are assumed for these sources.

We will not consider the contributions from intermediate-mass 
($\sim 1$--$8\,M_\odot$) stars in the following discussion. Such stars
are thought to be an important source of C in the Galaxy on timescales 
of $\gtrsim 1$ Gyr, thereby contributing to the increase of [C/O] in
Galactic stars from $\sim -0.5$ at [O/H]~$\lesssim -1$ to $\sim 0$ at 
[O/H]~$\sim 0$ (Akerman et al. 2004). Clearly, the delayed C contributions 
from intermediate-mass stars can be ignored for enrichment of the IGM
at early epochs ($z\gtrsim 15$ or $\lesssim 0.26$ Gyr since the big bang).
The precise epoch where intermediate-mass stars contribute to the IGM
is not known. However, there is no evidence for evolution of either
[C/H]$_{\rm IGM}$ or [Si/C]$_{\rm IGM}$ over at least the range of
$z=1.8$--4.1 (Schaye et al. 2003; Aguirre et al. 2004).
Therefore, we will only consider VMSs and SNe II as the possible sources
for metals in the IGM. As intermediate-mass stars produce C but 
no Si, the requirement of VMS contributions to account for the rather 
high [Si/C]$_{\rm IGM}$ will be strengthened by any possible contributions
from intermediate-mass stars.

\section{Metal Ejection from Low-Mass Halos}
We first explore the possibility that the metal inventory in
the IGM is the result of astration (i.e., star formation) in low-mass
($\sim 10^5$--$10^7\,M_\odot$) halos at $z\sim 15$--30. It is
widely recognized that the first stars were formed in low-mass halos
where gas could cool by H$_2$ molecules when the virial temperature
$T_{\rm vir}$ reached $\sim 300$ K (e.g., Couchman \& Rees 1986).
Haiman, Rees, \& Loeb (1997) showed that the soft UVB
produced by the first massive stars could dissociate H$_2$ molecules in
low-mass halos universally before the IGM was reionized.
Dissociation of H$_2$ molecules drastically changes the
condition for onset of astration to $T_{\rm vir}\sim 10^4$ K as required
for cooling by atomic species. This condition then governs astration for all
subsequent epochs including that of reionization. Ciardi, Ferrara,
\& Abel (2000) showed that universal H$_2$ dissociation
has not yet taken place for $z\sim 20$. While the actual domain of
astration by H$_2$ cooling remains to be established,
we consider the regime of $z\sim 15$--30 to be within
current estimates. Our purpose is to explore
if such astration could possibly provide
the metals in the IGM and what metal production rates are
required to achieve this.

The nature of the first stars made of big bang debris is not known.
They may be VMSs (see Abel, Bryan, \& Norman 2003; Bromm \& Larson 2004
for reviews). On the other hand, the discovery of a low-mass star with
[Fe/H]~$=-5.3$ (Christlieb et al. 2002)
suggests that low-mass stars may have formed very
early (e.g., Schneider et al. 2003)
and presumably a broader spectrum of normal stars including
SN II progenitors might be expected for the first stars.
In either case, the occurrence of a VMS or SN II in a low-mass
halo would disrupt the
baryonic gas in the halo and disperse it into the IGM on some scale.
To estimate the range of low-mass halos with gas expulsion, we use
models of hierarchical structure formation in the standard cold dark
matter cosmology as discussed in Barkana \& Loeb (2001).

A critical parameter for halo evolution is the virial temperature
$T_{\rm vir}$ of the gas. For $z>1$, $T_{\rm vir}$ is related to
the halo mass $M$ (mostly in dark matter) as
\begin{equation}
T_{\rm vir}\approx 211\left(\frac{\mu}{1.22}\right)
\left(\frac{M}{10^5\,M_\odot}\right)^{2/3}
\left(\frac{1+z}{10}\right)\ {\rm K},
\label{tvir}
\end{equation}
where $\mu$ is the mean atomic weight and $\mu=1.22$ or 0.6 for a
neutral or ionized gas, respectively,
with a primordial composition of H and He. We assume
that astration starts when the gas reaches a threshold virial
temperature $T_{\rm vir,0}$ for cooling. For H$_2$ cooling
we will take $T_{\rm vir,0}=300$ K ($\mu=1.22$).
Another critical parameter is the gas binding
energy $E_{\rm bi,gas}$. For $z>1$,
\begin{equation}
E_{\rm bi,gas}\approx 4.31\times 10^{47}
\left(\frac{M}{10^5\,M_\odot}\right)^{5/3}
\left(\frac{1+z}{10}\right)\ {\rm erg}.
\label{ebgas}
\end{equation}
We assume that all the gas would be expelled from the halo by an SN II
or VMS with an explosion energy $E_{\rm exp}$ if
$E_{\rm bi,gas}<E_{\rm exp}$ (e.g., Bromm, Yoshida, \& Hernquist 2003)
and retained for $E_{\rm bi,gas}\geq E_{\rm exp}$.
For illustration, we take
$E_{\rm exp}=10^{51}$ erg for SNe II and $4\times 10^{52}$ erg
for VMSs (HW).

For a halo associated with an $n\sigma$ density fluctuation (an
$n\sigma$ halo), the growth of its mass $M(z)$ as a function of
$z$ is prescribed by the Press-Schechter formalism (Press \& Schechter 1974;
see Fig. 6 in Barkana \& Loeb 2001).
Astration starts in a $3\sigma$ halo when the halo mass reaches
$M_0=3.6\times 10^4\,M_\odot$ at $z=26.9$. The gas in the halo would be
expelled by an SN II until the halo mass grows to
$M_{\rm bi}=7.0\times 10^6\,M_\odot$ at $z=18.6$ or by a VMS
until $M_{\rm bi}=7.1\times 10^7\,M_\odot$ at $z=15.3$.
In general, at a given $z$
an $n_0\sigma$ halo would reach $M(z)=M_0$ to start astration
and an $n_{\rm bi}\sigma$ halo would reach $M(z)=M_{\rm bi}$ to
greatly hinder gas expulsion. At any
given $z$ all $n\sigma$ halos with $n_0<n<n_{\rm bi}$ corresponding to
$M_0<M<M_{\rm bi}$ have started astration but will be disrupted by an
SN II or VMS. The relevant range of halos at $z=20$ corresponds to
$2.3<n<3.2$ ($5.5\times 10^4\,M_\odot<M<6.7\times 10^6\,M_\odot$)
for SNe II
or $2.3<n<3.8$ ($5.5\times 10^4\,M_\odot<M<6.1\times 10^7\,M_\odot$)
for VMSs.
For a given $z$, the fraction $F(M_0<M<M_{\rm bi}|z)$ of all matter in
$n\sigma$ halos with $n_0<n<n_{\rm bi}$ is
\begin{equation}
F(M_0<M<M_{\rm bi}|z)=\sqrt{\frac{2}{\pi}}\int_{n_0}^{n_{\rm bi}}
\exp\left(-\frac{x^2}{2}\right)dx.
\label{fmz}
\end{equation}
The function $F(M_0<M<M_{\rm bi}|z)$ for $T_{\rm vir,0}=300$ K
and $E_{\rm bi,gas}=10^{51}$ erg is very close to that for the same
$T_{\rm vir,0}$ but $E_{\rm bi,gas}=4\times 10^{52}$ erg as halos
with $E_{\rm bi,gas}>10^{51}$ erg are extremely rare for $z>15$.
A fraction $\sim 0.05$\% to 5\% of all matter is in
low-mass halos with gas expulsion at $z=30$ to 15.

We now estimate enrichment
of the IGM by gas expulsion from low-mass halos using O as a
representative element.
We consider a large reference region of the
universe and treat it as a closed homogeneous system. We assume
that the expelled gas is immediately
mixed with the entire IGM of the reference region. In this case,
the number ratio (O/H)$_{\rm IGM}$ of O to H atoms in the IGM as a
function of time $t$ (since the big bang) is determined by
\begin{equation}
\frac{d{\rm (O/H)}_{\rm IGM}}{dt}=
\sum_{n=n_0}^{n_{\rm bi}}P_{{\rm O},n\sigma}/{\rm (H)}_{\rm IGM},
\label{ohigm}
\end{equation}
where $P_{{\rm O},n\sigma}$ is the O production rate in an
$n\sigma$ halo and the sum extends over all the $n\sigma$ halos
with $n_0<n<n_{\rm bi}$ in the reference region. We explicitly
assume that $P_{{\rm O},n\sigma}$ is proportional to the number
(H)$_{n\sigma}$ of H atoms in the gas of the $n\sigma$ halo,
\begin{equation}
P_{{\rm O},n\sigma}=\Lambda_{\rm O}{\rm (O/H)}_\odot
{\rm (H)}_{n\sigma}.
\label{po}
\end{equation}
For $z\gg 1$, the majority of the H atoms in the reference region
reside in the IGM and the majority of the H atoms in a halo reside
in the gas. This gives
\begin{equation}
\frac{d({\rm O/H})_{\rm IGM}}{dt}\approx\Lambda_{\rm O}{\rm (O/H)}_\odot
F(M_0<M<M_{\rm bi}|z).
\label{ot}
\end{equation}
The constant parameter $\Lambda_{\rm O}$ is
related to the frequency of occurrence ($R$) for the O source
in a halo of mass $M$ as
\begin{equation}
R=\Lambda_{\rm O}X_{\rm O}^\odot
\frac{f_bM}{\langle Y_{\rm O}\rangle}=144\Lambda_{\rm O}
\left(\frac{M_\odot}{\langle Y_{\rm O}\rangle}\right)
\left(\frac{M}{10^5\,M_\odot}\right),
\label{freq}
\end{equation}
where $X_{\rm O}^\odot=9.6\times 10^{-3}$ is the solar mass fraction of O,
$f_b=0.15$ is the baryonic mass fraction for the halo, and
$\langle Y_{\rm O}\rangle$ is the average O yield of the source.
Equation (\ref{ot}) can be integrated to give
\begin{equation}
Z_{\rm O}^{\rm IGM}\equiv\frac{({\rm O/H})_{\rm IGM}}{{\rm (O/H)}_\odot}
=\Lambda_{\rm O}\int_0^{t(z)}F(M_0<M<M_{\rm bi}|z')dt',
\label{zo}
\end{equation}
where $t(z)=0.538[10/(1+z)]^{3/2}\ {\rm Gyr}$ for the
adopted cosmology and $t'\equiv t(z')$.
The above treatment is readily extended to mixtures from different sources.

The result for [O/H]$_{\rm IGM}=\log Z_{\rm O}^{\rm IGM}$ as a
function of $z$ is shown in Figure 1 for $\Lambda_{\rm O}=0.1\ {\rm Gyr}^{-1}$
using $T_{\rm vir,0}=300$ K for $M_0$ and $E_{\rm bi,gas}=10^{51}$ erg for
$M_{\rm bi}$ (solid curve).
Replacing the mass range $M_0<M<M_{\rm bi}$ with $M>10^5\,M_\odot$
gives essentially the same result (dot-dashed curve). The rate
$\Lambda_{\rm O}=0.1\ {\rm Gyr}^{-1}$ is representative of O
production by SNe II in the Galaxy [the solar O abundance is
produced by SNe II over $\sim 10$ Gyr]. For this rate [O/H]$_{\rm
IGM}=-3.5$ at $z=15$. So SNe II in low-mass halos cannot provide
the IGM inventory with a nominal abundance of [O/H]$_{\rm
IGM}=-2.3$ unless the relevant production rate is 16 times higher
than the Galactic rate. The requirement of $\Lambda_{\rm O}=1.6\
{\rm Gyr}^{-1}$ would also apply to any other sources (including VMSs) if
they were to produce the IGM inventory at early epochs.
Simulations by Norman, O'Shea, and Paschos (2004)
appear to suggest that VMSs could not enrich the IGM
to the level of [O/H]$_{\rm IGM}=-2.3$. However, this result
was based on putting one VMS in each halo with 
$M\geq 5\times 10^5\,M_\odot$ 
that was found in a simulation volume of 1 Mpc$^3$ at $z=15$. 
The number of such halos 
found in the simulation volume may not be statistically representative
(M. L. Norman, personal communication). Future numerical studies on
early enrichment of the IGM by VMSs or any other sources
should use larger simulation volumes to check the required metal production
rate against the value given here.

\section{Galactic Outflows}
We next consider contributions to the IGM from just galactic
outflows at $z<15$. We assume that only normal stars including SNe
II progenitors are formed and astration starts when $T_{\rm
vir,0}=10^4$ K ($\mu=0.6$) is reached for cooling by atomic
species. The halo mass at the onset of astration ranges from
$M_0=4.6\times 10^7\,M_\odot$ for $z=15$ to $2.6\times
10^8\,M_\odot$ for $z=4$. For these masses $E_{\rm
bi,gas}>10^{51}$ erg so an SN II cannot expel all the gas from the
halo. Subsequent to the onset of astration, we assume that a
fraction $\epsilon$ of the metals produced by SNe II is lost from
the halo and injected broadly into the IGM until the halo reaches
a cutoff mass $M_1$. While the dependence of $\epsilon$ on the
halo mass may be complicated, we will treat it as a constant to
determine how efficient mass loss must be to provide the IGM
inventory prior to $z=4$. Applying the same formalism leading to
equation (\ref{zo}), we have
\begin{equation}
Z_{\rm O}^{\rm IGM}=\epsilon\Lambda_{\rm O}
\int_{t_{15}}^{t(z)}F(M_0<M<M_1|z')dt',
\label{zout}
\end{equation}
where $t_{15}=0.26$ Gyr is the time corresponding to $z=15$.
Using $\Lambda_{\rm O}=0.1\ {\rm Gyr}^{-1}$ and $\epsilon=0.1$, we show
[O/H]$_{\rm IGM}=\log Z_{\rm O}^{\rm IGM}$ as a function of $z$
for $M_1=10^9$ and $10^{10}\,M_\odot$ (short and long dashed curves),
respectively, in Figure 1.
These $M_1$ values are consistent with observations of outflows from
Lyman break and dwarf galaxies (e.g., Pettini et al. 2001; Martin,
Kobulnicky, \& Heckman 2002). Values of $M_1>10^{10}\,M_\odot$ give
almost the same result as $M_1=10^{10}\,M_\odot$. Figure 1 gives
[O/H]$_{\rm IGM}=-2.9$ at $z=4$ for $M_1=10^{10}\,M_\odot$. Thus to achieve
[O/H]$_{\rm IGM}=-2.3$ at $z=4$ by just galactic outflows requires a rather
high but still reasonable $\epsilon=0.4$ for a Galactic metal 
production rate of $\Lambda_{\rm O}=0.1\ {\rm Gyr}^{-1}$ or
an equivalent $\epsilon\Lambda_{\rm O}=0.04$ Gyr$^{-1}$. Obviously,
smaller values of $\epsilon\Lambda_{\rm O}$ are required if galactic 
outflows are to provide some significant portions but not all of
the IGM inventory (see \S4).

\section{Elemental Ratios in the IGM and Implications for the Sources}
The results for $Z_{\rm O}^{\rm IGM}$ in \S\S2 and 3 can be extended
to other elements using the relevant production rates.
For two elements E and E$'$,
\begin{equation}
\frac{Z_{\rm E}^{\rm IGM}}{Z_{\rm E'}^{\rm IGM}}=
\frac{\Lambda_{\rm E}}{\Lambda_{\rm E'}}=
\frac{\langle Y_{\rm E}\rangle}
{\langle Y_{\rm E'}\rangle}
\left(\frac{X_{\rm E'}^\odot}{X_{\rm E}^\odot}\right),
\label{ye}
\end{equation}
where
\begin{equation}
\langle Y_{\rm E}\rangle=\frac{\int_{m_l}^{m_u}
Y_{\rm E}(m)\phi(m)dm}{\int_{m_l}^{m_u}\phi(m)dm}.
\label{aye}
\end{equation}
In the above equation $m$ is the mass in units of $M_\odot$ for
the source with $m_l$ and $m_u$ being the lower and upper limits,
$Y_{\rm E}(m)$ is the yield of element E as a function of $m$, and
$\phi(m)=m^{-\beta}$ is the IMF. We assume
a Salpeter IMF ($\beta=2.35$) for SNe II. As there is no a priori
knowledge on the mass distribution of VMSs, a wide range of IMFs
with $\beta=2.35$, 10, and 20 are explored. This allows us
to examine a wide range of abundance ratios of Si and Fe
relative to C and O resulting from VMS sources.
The average yields
$\langle Y_{\rm E}\rangle$ for C, O, Si, and Fe and the
corresponding [C/O], [Si/C], and [O/Fe] are given in Table 2 for a
number of models of VMSs [Umeda \& Nomoto 2002 (UN); HW] and SNe
II [Woosley \& Weaver 1995 (WW); UN; Chieffi \& Limongi 2004 (CL)]
for the $\beta$ values considered. Clearly, the elemental ratio
E/E$'$ for a source only depends on those stars with finite yields
of E and E$'$. For this reason, we give the limiting masses
for metal production as $m_l$ and $m_u$ in Table 2. Although
some VMSs or SNe II may form outside this mass range, they do not
affect the calculation of the elemental ratios.

Table 2 shows that VMSs have a narrow range of [C/O]~$=-0.71$ to
$-0.50$ for a wide range of IMFs but their [Si/C] varies from
1.18--1.30 for a Salpeter IMF to 0.61--0.84 for an IMF with
$\beta=20$ that very strongly favors the lower masses.
This is because the C and O yields are nearly constant for
all contributing VMSs while the Si yield increases significantly
(the Fe yield increases steeply) with mass (UN; HW). In contrast,
models of zero-metallicity SNe II (WW; UN; CL) give narrow ranges
of [C/O]~$=-0.23$ to 0.02 and [Si/C]~$=0.13$--0.17. For SNe II
with very low to solar metallicities, WW give [C/O]~$=-0.42$ and
[Si/C]~$=0.42$--0.50 while CL give [C/O]~$=-0.2$ and
[Si/C]~$=0.18$--0.28. We consider that SNe II in the regime of 
low metallicities
are relevant to the problem at hand. In all cases, [Si/C] is
high for VMSs but low for SNe II with zero or low 
metallicities. It is the high value of [Si/C]$_{\rm IGM}$ that
indicates significant VMS contributions as noted by Aguirre et al.
(2004) (see also Schaerer 2002). As VMSs have 
[Si/C]~$\gtrsim$~[Si/C]$_{\rm IGM}$ while SNe II have 
[Si/C]~$<$~[Si/C]$_{\rm IGM}$, this supports the existence of
VMS contributions but was not discussed in the recent critique of
VMS models by Tumlinson, Venkatesan, and Shull (2004).

We first compare the elemental ratios from stellar models with
[C/O]$_{\rm IGM}=0$ and [Si/C]$_{\rm IGM}=1.45$
inferred from UVB model Q. Neither VMSs nor SNe II of any metallicity
can account for both [C/O]$_{\rm IGM}$ and [Si/C]$_{\rm IGM}$.
As [C/O] is close to [C/O]$_{\rm IGM}$ for SNe II of zero
metallicity (or all metallicities according to CL) and [Si/C] is close to
[Si/C]$_{\rm IGM}$ for VMSs with a Salpeter IMF,
one may consider a mixture from these two
sources. However, to approximately match [C/O]$_{\rm IGM}$ would require
that almost all of the C come from SNe II and this would give a
[Si/C] for the mixture essentially equal to the SN II value, which is off by
$>1$ dex from the IGM value. Similarly, matching [Si/C]$_{\rm IGM}$ would
require that almost all of the C come from VMSs, thus giving
a very low [C/O] for the mixture. So in all cases and for all the stellar
models there is a problem in accounting for the elemental ratios inferred
from UVB model Q.

We now consider UVB model QG with [C/O]$_{\rm IGM} = -0.5$ and
[Si/C]$_{\rm IGM} = 0.74$.  These ratios cannot be matched by VMSs with a 
Salpeter IMF, for which [Si/C]~$ \geq 1.18$. A match can be obtained for VMSs 
if an IMF very strongly favoring the lower masses is used (see models 2B, 1C,
and 2C in Table 2). It is also possible to obtain a match with both
[C/O]$_{\rm IGM}$ and [Si/C]$_{\rm IGM}$
for UVB model QG if we consider mixtures of VMS
and SN II contributions. As the largest [Si/C]~$ = 1.3$ for VMSs (model 1A) 
and the largest [Si/C]~$= 0.42$ for SNe II of low
metallicity (model 6), to match [Si/C]$_{\rm IGM} = 0.74$ requires a
fraction of at least $f_{\rm C}^{\rm VMS} = 0.16$ of the C to come from VMSs.
Below we discuss four representative cases, all of which match [C/O]$_{\rm IGM}$ 
and [Si/C]$_{\rm IGM}$ for UVB model QG to within 0.1 dex (see Table 3). 

In case I the IGM inventory was produced by just VMSs
(IMF with $\beta=20$) in low-mass halos at early epochs
($z\gtrsim 15$). A high metal production rate of 
$\Lambda_{\rm O}^{\rm VMS}=1.6$ Gyr$^{-1}$ 
is required to achieve [O/H]$_{\rm IGM}=-2.3$ (\S2). 
This corresponds to a frequency of occurrence
$R_{\rm VMS}=4(M/10^5\,M_\odot)$ Gyr$^{-1}$ (eq. [\ref{freq}] for
$\langle Y_{\rm O}^{\rm VMS}\rangle=54.3\,M_\odot$) so that every
halo of $\sim 10^5\,M_\odot$ would have formed $\sim 1$ VMS by
$z=15$ ($t_{15}=0.26$ Gyr). This rate is usually considered the
upper limit for VMS formation as the UV irradiation from a VMS
suppresses further astration in its hosting halo (e.g., Couchman
\& Rees 1986). Case I does not allow any significant contributions 
to the IGM from galactic outflows at later epochs ($z<15$).
This appears to be in conflict with observations of
outflows from Lyman break and dwarf galaxies (e.g., Pettini et al.
2001; Martin et al. 2002). It is also at odds with the attribution
of the bulk of the IGM inventory to galactic outflows as advocated
by many previous studies (e.g., Madau, Ferrara, \& Rees 2001;
Scannapieco, Ferrara, \& Madau 2002; Thacker, Scannapieco, \&
Davis 2002). In addition, Bromm \& Loeb (2003) suggested that
VMSs may form only for metallicities below [C/H]~$=-3.5\pm 0.1$ and 
[O/H]~$=-3.0\pm 0.2$. However, for case I VMSs are
required to provide [C/H]~$=-2.9$ and [O/H]~$=-2.3$, which are
much beyond the suggested termination point for their formation.

In case II the sources for metals in the IGM are a combination of
VMSs and zero-metallicity SNe II in low-mass halos at early epochs.
A single Salpeter IMF ($\beta=2.35$) is used for the mass range
covering both VMSs and SNe II.
To achieve [O/H]$_{\rm IGM}=-2.3$ requires 
$\Lambda_{\rm O}^{\rm VMS}=0.9$ Gyr$^{-1}$ and
$\Lambda_{\rm O}^{\rm SNII}=0.7$ Gyr$^{-1}$. The latter is 7 times 
greater than the Galactic value. Thus case II requires high metal 
production rates for both VMSs and zero-metallicity SNe II.
As in case I, case II does not allow significant contributions from 
galactic outflows and requires VMSs to provide
enrichments beyond the suggested termination point for their 
formation.

For cases III and IV we consider a combination of early
contributions from VMSs (IMF with $\beta=10$ for case III and
Salpeter IMF for case IV) and later galactic outflows
driven by low-metallicity SNe II. Case III is chosen to match 
[Si/C]$_{\rm IGM}=0.74$ exactly. The required rate of
$\Lambda_{\rm O}^{\rm VMS}=1.0$ Gyr$^{-1}$ is high but the
required galactic outflows with an efficiency of $\epsilon=0.1$
for a Galactic rate of $\Lambda_{\rm O}^{\rm SNII}=0.1$ Gyr$^{-1}$
(for $M_1=10^{10}\,M_\odot$, see \S3)
are quite reasonable. Case III also requires VMSs to provide
[C/H]~$=-3.0$ and [O/H]~$=-2.5$, which are in excess of the termination
metallicities for their formation as suggested by Bromm \& Loeb (2003).
Case IV was calculated using $f_{\rm C}^{\rm VMS}=0.15$, which
requires VMSs to provide metals only up to the suggested 
termination point, and hence, a much smaller rate of
$\Lambda_{\rm O}^{\rm VMS}=0.3$ Gyr$^{-1}$. The required
efficiency of galactic outflows is $\epsilon=0.3$
for a Galactic $\Lambda_{\rm O}^{\rm SNII}$.
If we instead had required an exact match to 
[Si/C]$_{\rm IGM}$ for case IV, then $f_{\rm C}^{\rm VMS}=0.23$ 
with no major difference in the results. 

The results in Table 3 show that VMSs must be significant
contributors to the IGM based on the [Si/C]$_{\rm IGM}$ value and
that either a pure VMS source with an extreme IMF or
a blend of VMS and SN II sources
could provide the observed metal inventory in the IGM.
However, case IV appears to be the best scenario if we require
(1) major contributions from galactic outflows, (2) a Galactic 
$\Lambda_{\rm O}^{\rm SNII}$, (3) a nonextreme IMF for VMSs, and
(4) strict termination metallicities for VMS formation as suggested 
by Bromm \& Loeb (2003). In case IV
VMSs in low-mass halos provided 15\% of the C, 20\% of the O,
and 51\% of the Si prior to H$_2$ dissociation and reionization, and
the rest of the IGM inventory was provided by
later efficient outflows driven by SNe II in intermediate-mass halos.
This suggests that VMSs must be substantial 
contributors to the IGM
with high frequencies of occurrences at very high $z$.
Venkatesan \& Truran (2003) calculated that VMSs would provide
10 H-ionizing photons per baryon in order to reach 
[O/H]$_{\rm IGM}=-2.5$. So the VMSs in case IV would provide 
$\approx 3$ H-ionizing photons per baryon in reaching
[O/H]~$=-3.0$ (using C gives the
same result. Note that the value of 0.35 H-ionizing photons per baryon
given in Venkatesan \& Truran 2003 assumes a termination metallicity
of [O/H]~$\sim -4.0$ for VMS formation). 
This may be consistent with the ``weak'' version of
the VMS scenario discussed in Tumlinson et al. (2004).

So far our discussion has focused on C, O, and Si as these elements
are measured in the IGM. The abundances of other elements such as Fe
can also be calculated for the four cases discussed above. The O and Fe
yields in Table 2 give [O/Fe]~$=0.71$ and 0.52 for cases I and III,
respectively, and [O/Fe]~$=0.12$ for cases II and IV (see Table 3).
For [O/H]$_{\rm IGM}=-2.3$, this implies [Fe/H]$_{\rm IGM}= -3.0$ and 
$-2.8$ for cases I and III, respectively, and [Fe/H]$_{\rm IGM}= -2.4$
for cases II and IV. Observations of metal-poor stars in the Galaxy 
(e.g., Cayrel et al. 2004)
show that [O/Fe]~$\gtrsim 0.5$ for [Fe/H]~$\sim -3.0$ to $-2.4$.
Cases I and III are consistent with these observations. However, 
the [O/Fe] value for cases II and IV is too low compared with
those observed in metal-poor stars.
This may be a potentially important problem for these two cases, 
especially for case IV, which satisfies all the other criteria
discussed above.
The low [O/Fe] value for cases II and IV is mostly
due to the low values of [O/Fe]~$\sim 0.2$ given by SN II models. 
We note that the Fe yields of SNe II were not calculated 
a priori but generated by employing largely artificial mass cuts 
in all models. Stellar observations (e.g., Nissen et al. 2002)
show that [O/Fe]~$\sim 0.5$ for the regime of 
$-2.4<{\rm [Fe/H]}<-1.5$, where SNe II are clearly the dominant
source for both O and Fe. This indicate that the calculated
average Fe yields of SNe II are a factor of $\sim 2$ too high.
Much work is required to resolve the problem of Fe yields and the 
associated uncertainties in [O/Fe] for SN II models. We also note
that our next best case III gives an adequate [O/Fe]
but requires VMSs to provide [C/H] and [O/H] somewhat
beyond the termination
metallicities suggested by Bromm \& Loeb (2003). If the criteria
for termination of VMS formation could be relaxed, then case III
may be a viable alternative to case IV in accounting for 
the IGM data and stellar observations.

Clearly, the results presented here are subject to uncertainties in 
theoretical models of stellar yields and in IGM
measurements and analyses.
Possible uncertainties are unlikely to explain the large discrepancies
between stellar models and the elemental ratios from UVB model Q but may
affect some of our results regarding model QG. As discussed above,
uncertainties in Fe yields of SN II models affect the [O/Fe] values
for cases II and IV. In addition, the
required rates quoted above (see Table 3)
correspond to a nominal abundance of
[O/H]$_{\rm IGM}=-2.3$ and would change if different values were used.
For this nominal value all four cases considered (including case I for 
a pure VMS source) imply [Fe/H]$_{\rm IGM}\geq -3.0$ (using
[O/Fe]~$=0.5$ instead of 0.12 for cases II and IV does not change
the result). This is high considering a number of
ultra-metal-poor stars in the Galaxy have $-4\lesssim$~[Fe/H]~$<-3$
(e.g., Christlieb 2003). One possible explanation would be that those 
ultra-metal-poor stars formed in small halos in under-enriched 
regions and were later accreted into the Galaxy. This is consistent 
with the large scatter for the lognormal distribution of metal
abundances in the IGM and with the standard model of hierarchical
structure formation.
The above issues and the general problem of relating abundances
in the IGM to those in
metal-poor stars remain to be addressed.

In summary, we find that VMSs must be significant contributors to the
IGM. Our preferred scenario is that VMSs provided between 15\%
(case IV) and 60\% (case III) of the C (and corresponding levels of
other metals) in the IGM by disrupting the
gas in low-mass ($\sim 10^5$--$10^7\,M_\odot$) halos at very high 
redshift ($z\gtrsim 15$). Outflows driven by SNe II in 
intermediate-mass ($\sim 10^8$--$10^{10}\,M_\odot$) halos provided
the remainder (the bulk in case IV)
of the metals during later epochs 
($4<z\lesssim 6$, see Fig. 1). Thus, there is a large gap in $z$
between metal production by VMSs very early in the chemical evolution 
of the IGM and subsequent contributions from galactic outflows.
The metals from these two different sources appear to be well
mixed locally but have a wide range in net abundances as established
by Schaye et al. (2003), Aguirre et al. (2004), and Simcoe et al. (2004).

\acknowledgments
We thank the referee, Aparna Venkatesan, for a thorough report
and many helpful suggestions that greatly improve the paper.
This work was supported in part by DOE grants DE-FG02-87ER40328
(Y. Z. Q.) and DE-FG03-88ER13851 (G. J. W.),
Caltech Division Contribution 9010(1115).

\clearpage
\begin{deluxetable}{lrrr}
\tablecolumns{4}
\tablewidth{0pc}
\tablecaption{IGM Data\tablenotemark{a}}
\tablehead{
\colhead{UVB Model}&\colhead{[O/H]$_{\rm IGM}$}&\colhead{[C/O]$_{\rm IGM}$}
&\colhead{[Si/C]$_{\rm IGM}$}}
\startdata
Q&$-2.3$ to $-2.0$&0&1.45\\
QG&$-2.3$ to $-2.0$&$-0.5$&0.74\\
\enddata
\tablenotetext{a}{[C/O]$_{\rm IGM}$ from Simcoe et al. 2004,
[Si/C]$_{\rm IGM}$ from Aguirre et al. 2004, and [O/H]$_{\rm IGM}$ from
Simcoe et al. 2004 or converted from [C/H]$_{\rm IGM}$ of
Schaye et al. 2003 using [C/O]$_{\rm IGM}$. Solar abundances adopted:
$\log\epsilon_\odot({\rm C})=\log({\rm C/H})_\odot+12=8.52$,
$\log\epsilon_\odot({\rm O})=8.83$ (Grevesse \& Sauval 1998), and
$\log\epsilon_\odot({\rm Si})=7.55$ (Anders \& Grevesse 1989).}
\end{deluxetable}

\begin{deluxetable}{lrrrrrrrrrr}
\tabletypesize{\footnotesize}
\tablecolumns{11}
\tablewidth{0pc}
\tablecaption{VMS and SN II Yields and Yield Ratios}
\tablehead{
\colhead{}&\colhead{Metallicity}&\colhead{}&\colhead{}&
\multicolumn{4}{c}{$\langle Y_{\rm E}\rangle$ ($M_\odot$)}&\colhead{}&
\colhead{}&\colhead{}\\
\cline{5-8}\\
\colhead{Model}&\colhead{$(Z/Z_\odot)$}&\colhead{($m_l,m_u$)}&
\colhead{$\beta$}&\colhead{C}&\colhead{O}&\colhead{Si}&\colhead{Fe}&
\colhead{[C/O]\tablenotemark{a}}&\colhead{[Si/C]\tablenotemark{a}}&
\colhead{[O/Fe]\tablenotemark{b}}}
\startdata
\sidehead{VMSs}
1A (UN)&0&(150, 270)&2.35&3.67&51.0&18.4&6.85&$-0.71$&1.30&0.00\\
2A (HW)&0&(140, 260)&2.35&4.34&44.2&16.6&6.21&$-0.57$&1.18&$-0.02$\\
1B (UN)&0&(150, 270)&10&4.56&52.4&12.7&2.99&$-0.62$&1.05&0.38\\
2B (HW)&0&(140, 260)&10&4.99&46.4&9.24&0.67&$-0.53$&0.87&0.97\\
1C (UN)&0&(150, 270)&20&5.41&54.3&9.41&1.42&$-0.57$&0.84&0.71\\
2C (HW)&0&(140, 260)&20&5.51&47.2&5.59&0.046&$-0.50$&0.61&2.14\\
\sidehead{SNe II}
3 (WW)&0&(12, 30)&2.35&0.12&0.30&0.039&0.099&0.02&0.13&$-0.38$\\
4 (UN)&0&(13, 30)&2.35&0.28&1.29&0.098&0.083&$-0.23$&0.15&0.32\\
5 (CL)&0&(13, 35)&2.35&0.31&1.19&0.11&0.10&$-0.15$&0.17&0.19\\
6 (WW)&0.01&(12, 40)&2.35&
0.20&1.43&0.13&0.13&$-0.42$&0.42&0.17\\
7 (CL)&0.005&(13, 35)&2.35&
0.34&1.49&0.13&0.10&$-0.21$&0.18&0.28\\
8 (WW)&1&(11, 40)&2.35&0.19&1.37&0.15&0.12&$-0.42$&0.50&0.20\\
9 (CL)&1&(15, 35)&2.35&0.36&1.54&0.17&0.14&$-0.20$&0.28&0.19\\
\enddata
\tablenotetext{a}{Calculated using the same solar abundances as in Table 1.}
\tablenotetext{b}{Calculated using $X_{\rm O}^\odot=9.6\times 10^{-3}$
and $X_{\rm Fe}^\odot=1.3\times 10^{-3}$ (Anders \& Grevesse 1989).}
\end{deluxetable}

\clearpage
\begin{deluxetable}{cccrrrrrrrr}
\tablecolumns{10}
\tablewidth{0pc}
\tablecaption{Representative Cases to Match the IGM Data
from UVB Model QG}
\tablehead{
\colhead{}&\colhead{VMS}&\colhead{SN II}&\colhead{}&\colhead{}&\colhead{}&
\colhead{}&\colhead{}&\colhead{$\Lambda_{\rm O}^{\rm VMS}$}&
\colhead{$\Lambda_{\rm O}^{\rm SNII}$}&\colhead{}\\
\colhead{Case}&\colhead{Model}&\colhead{Model}&
\colhead{$f_{\rm C}^{\rm VMS}$\tablenotemark{a}}&
\colhead{[C/O]}&\colhead{[Si/C]}&
\colhead{$f_{\rm O}^{\rm VMS}$\tablenotemark{a}}&\colhead{[O/Fe]}&
\colhead{(Gyr$^{-1}$)\tablenotemark{b}}&
\colhead{(Gyr$^{-1}$)\tablenotemark{b}}&
\colhead{$\epsilon$\tablenotemark{b}}}
\startdata
I&1C&\nodata&1&$-0.57$&0.84&1&0.71&1.6&\nodata&\nodata\\
II&1A&4&0.28&$-0.42$&0.82&0.54&0.12&0.9&0.7&\nodata\\
III&2B&6&0.60&$-0.49$&0.74&0.66&0.52&1.0&0.1&0.1\\
IV&2A&6&0.15&$-0.45$&0.65&0.20&0.12&0.3&0.1&0.3\\
\enddata
\tablenotetext{a}{Specified by the same Salpeter IMF governing both
VMSs and SNe II for case II, chosen to match [Si/C]~$=-0.74$ exactly
for case III,
and chosen to obtain [O/H]~$=-3.0$ from VMSs for case IV.}
\tablenotetext{b}{Rates and efficiencies
required to provide [O/H]~$=-2.3$.
The required rate $\Lambda_{\rm O}^{\rm VMS}$ for VMSs in all cases
and the required rate $\Lambda_{\rm O}^{\rm SNII}$ for
zero-metallicity SNe II in case II are given assuming
complete disruption of the gas by these sources in low-mass halos.
For galactic outflows in cases III and IV, the required outflow
efficiency $\epsilon$ is given assuming a Galactic 
$\Lambda_{\rm O}^{\rm SNII}$ (an equivalent 
$\epsilon\Lambda_{\rm O}^{\rm SNII}$ also satisfies the requirement).}
\end{deluxetable}

\clearpage
\figcaption{Evolution of [O/H]$_{\rm IGM}$ as a function of $z$ for
gas expulsion from low-mass halos with astration controlled
by H$_2$ cooling (solid curve for contributing halos with
$M_0<M<M_{\rm bi}$ determined by $T_{\rm vir,0}=300$ K for onset of
astration and $E_{\rm bi,gas}=10^{51}$ erg for onset of gas retention, 
dot-dashed curve for 
contributing halos with $M>10^5\,M_\odot$) or outflows from
intermediate-mass halos subsequent to H$_2$ dissociation
($T_{\rm vir,0}=10^4$ K, short and long-dashed curves for cutoff
masses of $10^9$ and $10^{10}\,M_\odot$, respectively). The vertical
line at $z=15$ divides the two regimes. For both regimes the reference
production rate $\Lambda_{\rm O}=0.1$ Gyr$^{-1}$ corresponds to the
Galactic rate for SNe II.}
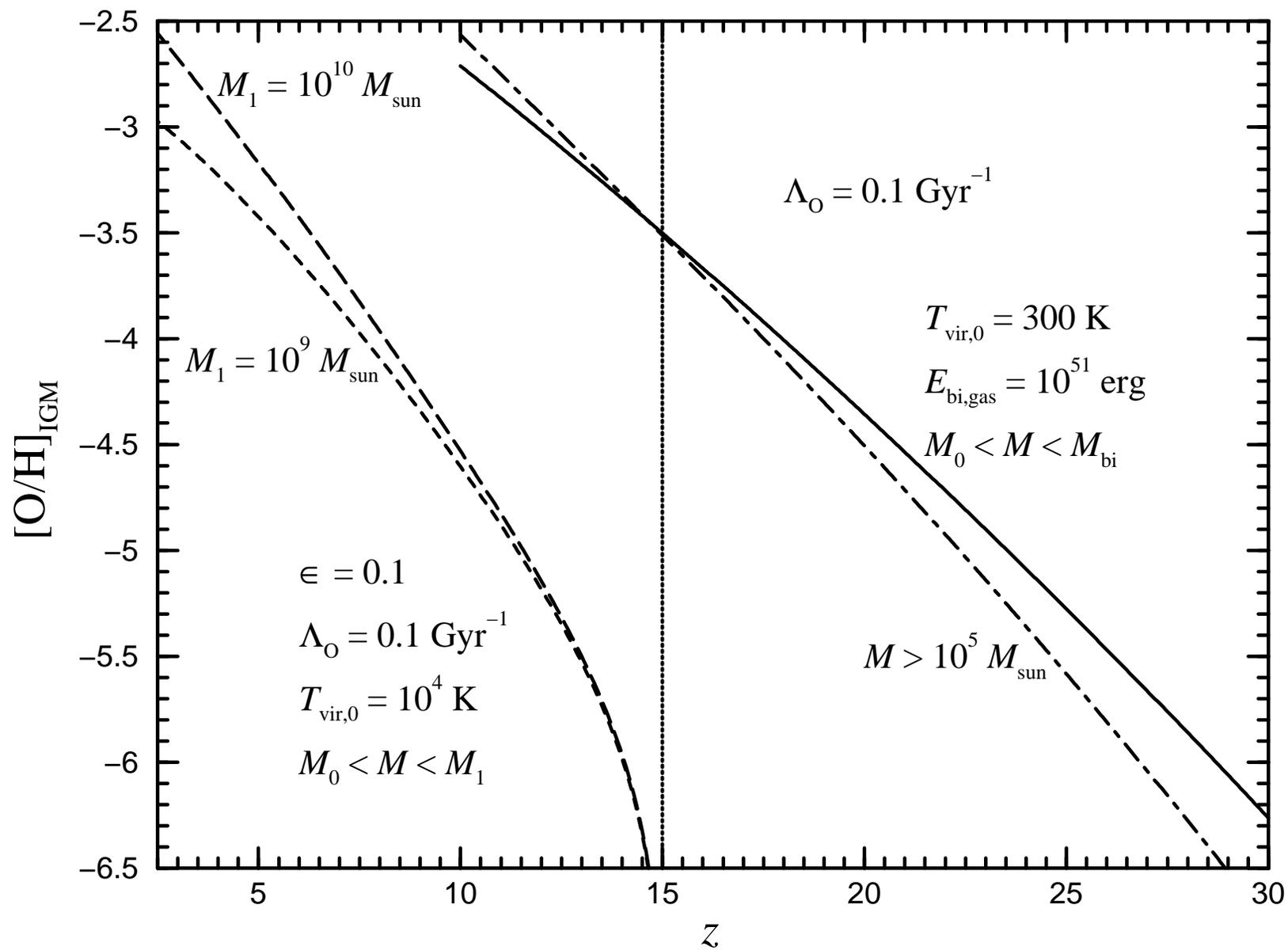
\end{document}